\documentclass[aps,prl,nofootinbib,floatfix,twocolumn,showpacs]{revtex4-1}

\usepackage{pstricks}
\usepackage{graphicx}
\usepackage{bm}
\usepackage{amsmath,amssymb,amsfonts}

\usepackage[sort&compress]{natbib}

\begin{document}

\title{Probing Positron Gravitation at HERA}

\author{Vahagn Gharibyan}
\email[]{vahagn.gharibyan@desy.de}
\thanks{I thank B.~Sobloher and S.~Schmitt for providing details about the
positron generated spectra, and R.~Brinkmann for details about the electron 
measurement and the HERA. I'm thankful also to 
 A.~Buniatyan and K.~Balewski for useful discussions.}
\affiliation{Deutsches Elektronen-Synchrotron DESY - D-22603 Hamburg}

\begin{abstract}

An equality of particle and antiparticle gravitational interactions
holds in general relativity and is supported by indirect 
observations.
Here I develop a method based on high energy Compton scattering
to measure the gravitational interaction of accelerated charged particles.
Within that formalism the Compton spectra measured at HERA rule out 
the positron's anti-gravity and hint for a positron's 1.3(0.2)\% weaker
coupling to the gravitational field relative to an electron. 
\end{abstract}


\maketitle

\section{Introduction}

The weakness of gravitation makes it the least experimentally
investigated interaction among the fundamental forces of
nature. The weak gravity combined with a rarity and 
vulnerability of antiparticles drives any attempt of testing
the antimatter gravitation to its technical limits.
Einstein's general  
relativity~\cite{Einstein-GR}, the currently accepted 
theory of gravitation, does not distinguish
between particles and antiparticles.
Strictly speaking, general relativity deals only with masses 
and light, leaving gravitational interaction of particles, 
antiparticles, or photons to be established in
quantum gravity theory~\cite{AmelinoCamelia:2008qg}. 
Hence, observations of antiparticle gravitation could serve as an 
experimental input for quantum gravity~\cite{Nieto:1991xq}.
Additional motivations for such investigation are the
still unexplained matter-dominant universe~\cite{Agashe:2014kda} and the
connection of antimatter's possible anti-gravity~\cite{Villata:2011bx} 
to the accelerated expansion of the universe~\cite{Riess:1998cb}. 
One can also think about a possible particle-antiparticle gravitational 
asymmetry from an analogy to electroweak interactions, where a
photon's massive partners, W and Z bosons, are considered responsible
for space and charge parity violations~\cite{Beringer:1900zz}. 
Thus, possible massive or lower spin gravitons could introduce similar 
violations~\cite{Goldhaber:2008xy} that may remain hidden at low energies 
and will become detectable at high energies.

Indirect observations of matter-antimatter gravitational asymmetry 
involve nuclei with different content of quark-antiquarks  
in the equivalence principle E$\ddot{o}$tv$\ddot{o}$s type 
experiments~\cite{Schlamminger:2007ht,Adelberger:2009zz}.  
Using $CPT$ conservation the observed stringent limits for the 
equivalence principle violating matter could be expanded to 
a limit below $10^{-7}$ for the matter-antimatter low energy 
gravitational asymmetry~\cite{Alves:2009jx}.  
Technical difficulties for  charged antiparticle's 
gravitational coupling direct measurements
turned physicists' attention to  neutral antimatter 
tests~\cite{Scampoli:2014tpa,  Amole:2013gma, Gabrielse:2012xe} 
which may  deliver conclusive results soon.
The ongoing experiments, however, are still at low energy, and massive
gravitons' interactions may remain unseen.

In this Letter I will demonstrate an extreme sensitivity of a high energy 
process - laser Compton scattering - to an antiparticle's  hypothetical
anti-gravity and gravitational charge parity violation.
Next, applying the developed formalism to the 
existing data of the HERA Compton polarimeter, I will compare
the $\gamma$-spectra  generated by 
electrons and positrons to measure the charge asymmetry 
for their gravitational interaction.
Systematic effects and prospects for other tests will be discussed at 
the end.

\section{Gravity introduced dispersion}

In an earlier publication,  high energy Compton scattering 
sensitivity has been shown to a Planck scale refractive and birefringent vacuum 
model~\cite{Gharibyan:2012gp}. 
Subsequently, I applied the same formalism to the Earth's gravity
assuming the real gravitational field induced dispersion only for the
Compton photons~\cite{Gharibyan:2014mka}. 
The dispersion, however,  also affects the 
leptons  involved in the scattering~\cite{Evans:2001hy} in agreement 
with the equivalence principle.
This makes the ref.~\cite{Gharibyan:2014mka} conclusions about 
the general relativity violation invalid~\cite{khaladgyan}.

Here I follow the formalism developed by Evans et al.~\cite{Evans:2001hy} 
to find a massive particle's energy-momentum or dispersion relation in
a static and isotropic gravitational field described by the Schwarzschild metric.
Combining the Eq.(3) and Eq.(30) from the reference \cite{Evans:2001hy}, 
for the Earth's weak field, one can derive a dispersion relation 
\begin{equation}
\frac{P}{\cal E} = \beta + \frac{2 G M_\oplus}{ R_\oplus},
\label{disp}
\end{equation}
where $G$ is the gravitational constant and 
${\cal E},P,\beta$ are  energy, momentum, velocity of the particle 
($c=1$ is assumed throughout the Letter).
This relation is also valid  for massless particles. Indeed, at
$\beta=1$ it describes the photon refraction in a gravitational field
in a form derived by many authors; see ref.~\cite{de Felice:1971ui} 
and references therein, or for a more recent reference, see ref.~\cite{Sen:2010zzf}.

To allow departure from the equivalence principle let us 
retain the interaction strength $G$ for matter particles 
and use a different strength $G_p$ for antimatter leptons  
to write  Eq.(\ref{disp}) for positrons in the 
following form
\begin{equation}
\frac{P}{\cal E} = \beta + \frac{2 G M_\oplus}{ R_\oplus}\biggl(1+\frac{\Delta G}{G}\biggr),
\label{disp1}
\end{equation}
with \hbox{$\Delta G = G_p - G$}.
For an anti-gravitating positron \hbox{$G_p=-G$}.

\section{The Compton process affected by gravity}

Using energy-momentum conservation with  Eq.(\ref{disp}) and  Eq.(\ref{disp1}), 
when in the Earth's gravitational field a photon scatters off a 
positron with energy ${\cal E}$, 
the Compton scattering kinematics is given by
\begin{equation}
{\cal E}x - \omega (1+x+\gamma^2 \theta^2) + 
4\omega \biggl(1 - \frac{\omega}{{\cal E}} \biggr) \gamma^2 
\frac{ M_\oplus }{ R_\oplus}\Delta G = 0,
\label{comp0}
\end{equation}
where \hbox{$x=4\gamma \omega_0\sin^2{(\theta_0/2)}/m$}, 
with $\gamma$ and $m$ being the Lorentz factor and mass of the
initial positron, respectively. The initial photon's energy and angle are denoted
by $\omega_0$ and $\theta_0$, while the dispersion of  Eq.(\ref{disp}) is in effect 
for the
scattered photon with energy $\omega$ and angle $\theta$; the angles are defined 
relative to the initial positron. This kinematic
expression is derived for weak gravity and high energies,  
i.e., the  $\mathcal{O}((GM_\oplus/R_\oplus)^2)$, $\mathcal{O}(\theta^3)$, and 
$\mathcal{O}(\gamma^{-3})$ terms are neglected.
To determine the outgoing photon's maximal energy,  Eq.(\ref{comp0}) is solved 
for $\omega$ at $\theta=0$ with the following result:
\begin{equation}
\omega_{max} = {\cal E}~\frac {b+q-\sqrt{b^2+q(q-2b+4)} } {2\,q},
\label{comp}
\end{equation}
where $b=1+x$ and $q=2\gamma^2 M_\oplus \Delta G / R_\oplus$.
Thus, in high energy Compton scattering
the factor $\Delta G$ is amplified by $\gamma^2$, 
allowing one to measure it by detecting the extreme energy of the scattered photons 
$\omega_{max}$, or positrons \hbox{${\cal E}-\omega_{max}$} (Compton edge).  

\begin{figure}[h]
\centering
\includegraphics[scale=0.47]{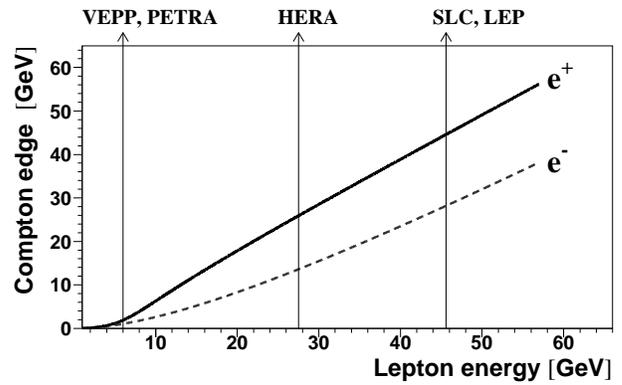}
\caption{\label{fig1}
The maximal energy of Compton scattered photons (Compton edge)
and its
dependence on the initial lepton energy 
for a head-on collision with 532nm laser light.
Solid and dotted lines correspond to 
matter gravity (electron, $G$) and antimatter
anti-gravity  (positron, -G)  respectively. 
Names of $e^+ e^-$ accelerators are printed at the upper part.}
\end{figure}

In order to estimate the method's sensitivity, I calculate the Compton edge
for an incident photon energy 2.32~eV (the widely popular green laser) 
at different energies of the accelerator leptons.  
The resulting dependencies for a matter (electron) gravity and antimatter 
(positron) anti-gravity are presented in Fig.~\ref{fig1}.
The plot shows considerable sensitivity, which grows toward high energies
in a range available to  accelerating laboratories.
For handling  measurement's systematic errors, from an experimental point of view,
it is more precise to measure a relative asymmetry rather than absolute Compton 
edge energy. 
Therefore, we form an asymmetry of Compton edges measured on positrons
($\omega_{max}^p$) and electrons ($\omega_{max}^e$)
\begin{equation}
A=\frac{\omega_{max}^p-\omega_{max}^e}{\omega_{max}^p+\omega_{max}^e}
\label{asym}
\end{equation}
and use Eq.(\ref{comp0}) to find the charge parity gravitational violation 
magnitude
\begin{equation}
\frac{\Delta G}{G}=\frac{2A(1-A)(1+x)^2}{(1+A)(2Ax+A-1)}
\biggl(4 \gamma^2 \frac{G M_\oplus}{ R_\oplus} \biggr)^{-1}.
\label{dgog}
\end{equation}

\section{Experimental results}

\begin{figure}[h]
\centering
\includegraphics[scale=0.49]{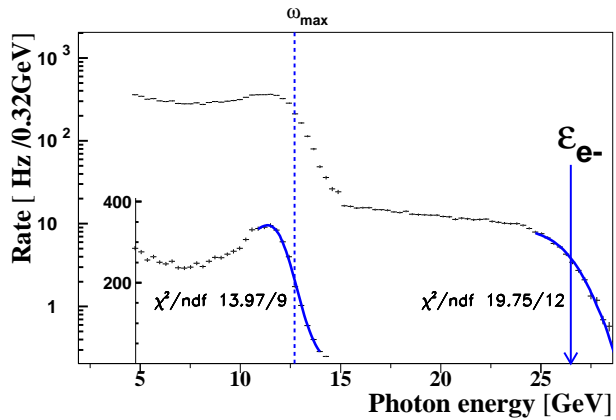}
\caption{\label{fig2}
HERA polarimeter Compton $\gamma$-spectrum 
produced by laser scattering on $26.5~GeV$ electrons,
on top of background Bremsstrahlung with fit results.
The inset displays the background subtracted Compton spectrum.
Vertical  lines show measured values of  the Compton ($\omega_{max}$) 
and Bremsstrahlung (${\cal E}_{e-}$) 
maximal energies. }
\end{figure}

The high-energy accelerators where laser Compton facilities have been operated 
for years, are listed on the upper energy scale of  Fig.~\ref{fig1}. 
As can be seen from the plot, 6~GeV storage rings  have low 
sensitivity while the higher energy colliders (HERA, SLC, LEP) have a great
potential for detecting gravity related energy shifts.
This is true for the HERA and SLC Compton polarimeters but not for the LEP 
polarimeter, which has
generated and registered many photons per machine pulse~\cite{LEP-polarimeter}. 
In this multi-photon regime, any shift of the Compton edge is convoluted with the 
laser-electron luminosity and can-not be disentangled and measured separately.
Unlike the LEP, the SLC polarimeter  operated in multi-electron mode and  
analyzed the energies of interacted leptons using a magnetic 
spectrometer~\cite{ALEPH:2005ab}. However, at SLC only the electron beam was polarized, 
and positron data are missing. Hence, we turn to HERA, which have recorded
Compton measurements for both the electrons and the positrons.
At the HERA transverse polarimeter Compton photons are registered by a calorimeter 
in single particle counting mode. A recorded Compton spectrum 
produced by $514.5$nm laser scattering on $26.5~GeV$ electrons,
 from ref.\cite{Barber:1992fc}, 
is shown in Fig.~\ref{fig2} superimposed on a background  Bremsstrahlung distribution.
In contrast to  Compton scattering, in the Bremsstrahlung process the momentum 
transfer is not fixed, and any small dispersive  effect is smeared out and becomes 
negligible~\cite{Gharibyan:2003fe}.
Hence, following the analysis in  ref.~\cite{Gharibyan:2003fe}, I calibrate the energy 
scale according to the maximal Bremsstrahlung energy which is found by fitting 
a convolution of parent energy distribution ${d\Sigma}/{d\omega}$ with the 
detector response gaussian function,
\begin{equation}
F(E_\gamma)=N\int^{E_m}_{0} 
\frac{d\Sigma}{d\omega}\frac{1}{\sqrt{\omega}} 
\exp\Biggl({\frac{-(\omega-E_\gamma)^2}
{2\sigma_0^2 \omega}\Biggr) d\omega},
\label{eqfold}
\end{equation}
to the Bremsstrahlung spectrum.
$\sigma_0$  and $E_\gamma$ in the fitting function denote the
calorimeter resolution and detected photon's energy respectively
while the normalizing factor  $N$ and maximal energy $E_m$ are 
free fitting parameters.
The same fitting function with the Bremsstrahlung parent distribution
replaced by the Compton scattering differential cross-section 
${d\Sigma_C}/{d\omega}$ 
is applied to the background subtracted spectrum to find the Compton
edge  at \hbox{$\omega_{max}^e= 12.70\pm 0.02$}~GeV. The fit results together
with fit quality estimates are shown in  Fig.~\ref{fig2}.
More details about the analysis and experimental setup can be found 
in the ref.~\cite{Gharibyan:2003fe}.

\begin{figure}[h]
\centering
\includegraphics[scale=0.49]{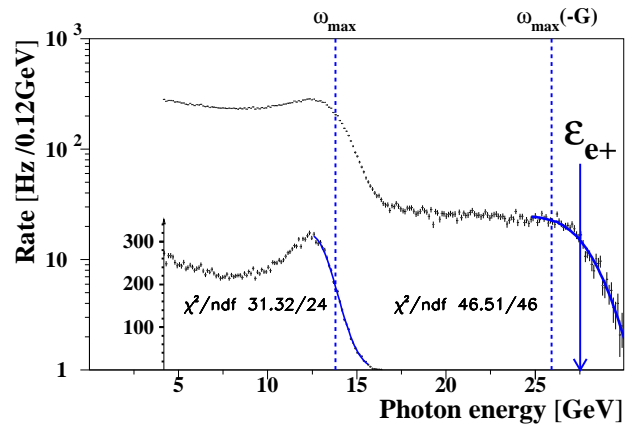}
\caption{\label{fig3}
A similar plot to Fig.~\ref{fig2} for positrons with energy $27.5~GeV$. 
The Compton edge energy for anti-gravitating positrons is indicated by
a vertical line $\omega_{max}(-G)$.}
\end{figure}

The same analysis procedure is applied to a HERA polarimeter 
Compton spectrum that was generated with $27.5~GeV$ positrons and 
has been reproduced in Fig.~8 of ref.~\cite{Sobloher:2012rc}. The resulting 
plots with fit quality outcomes are displayed in Fig.~\ref{fig3}.
Comparing the obtained Compton edge  
\hbox{$\omega_{max}^p= 13.80\pm 0.02$}~GeV 
with the photons' maximal energy for the anti-gravitating positrons 
$25.9$~GeV, derived from  Eq.(\ref{comp}),  
one can conclude without any advanced systematic error analysis
that anti-gravity for the positrons is ruled out.

Since the spectra for electrons and positrons are detected with the
same experimental setup, i.e. with the same laser, geometry and detector,
both measurements will experience the same systematic influences 
that will cancel out or reduce greatly  in the asymmetry of Eq.(\ref{asym}). 
Hence, we omit systematic corrections or errors described in the 
refs.~\cite{Barber:1992fc,Gharibyan:2003fe,Sobloher:2012rc}  
and use only the quoted statistical errors
for the $\omega_{max}^e$, $\omega_{max}^p$ to obtain a
positron-electron Compton edge asymmetry 
\hbox{ $A = 0.01297 \pm 0.00147$}.

In order to account for the different energies
of accelerated electrons and positrons $26.5$ and $27.5$ GeV, 
the measured maximal energies in the asymmetry
calculation have been normalized to 
$13.10$ and $13.80$~GeV
for the electrons and positrons, respectively.
These are the expected Compton edge values from  Eq.(\ref{comp}) 
in the absence of gravitational anomaly, at $\Delta G = 0$.
Normalization uncertainty associated with the 
laser and lepton beam energy spread is included 
in the asymmetry error.

Inserting the asymmetry into  Eq.(\ref{dgog}), we obtain a measured  
charge parity gravitational violation value
\begin{equation}
\frac{\Delta G}{G}=-0.0133\pm 0.0015,
\label{result}
\end{equation}
which differs from zero within a $9\sigma$ confidence.
The obtained negative sign corresponds to a weaker gravitational coupling
for the positrons relative to the electrons.

\section{Conclusions}

Applying a gravitational field-induced dispersion 
and assuming an equivalence principle violation in a general form 
$\Delta G/G$ for  positrons, an outstanding
sensitivity has been demonstrated for the high energy Compton scattering 
to such gravitational anomaly. 
Within the developed formalism, the HERA Compton polarimeter's 
recorded spectra with electrons and positrons strongly disfavor the positron's 
anti-gravity and show a significant deviation of the $\Delta G/G$ from zero.
The last claim is based on a detected $1.3\%$ energy asymmetry, which is a large 
number compared to the laser and lepton beam energy relative uncertainty of 
$10^{-5}$ and $10^{-3}$, respectively. The remaining  source of a
possible systematic energy error is the detector that is eliminated 
from final result by using the asymmetry instead of absolute energy measurements.  
However, additional uncorrelated systematic errors  may impair the outcome 
and, claiming a definite observation of charge parity violation at high energy 
gravitational interactions would require the following:

-- a thorough analysis of many Compton spectra 
accumulated and recorded by the HERA during its running period;

-- elimination of possible electroweak sources  that can mimic such result;

-- experimental verification at other accelerators.

\noindent In the absence of these, the measured electron-positron asymmetry could 
only be called a hint for the gravitational symmetry breaking and an invitation 
for further studies. 
New experiments, however, will require future $e^-e^+$ machines with sufficiently 
high $\gamma$ or a precise setup on the currently running 6~GeV accelerator 
PETRA-III with the highest positron energy available.
Anyway, it is worth the efforts since high energy violation of the equivalence
principle and gravitational charge parity 
could reveal an interaction to massive or lower spin gravitons with
a possible relation to dark matter or energy.

\end{document}